# A Novel Approach to Encourage Students' Independent Thinking in the Physics Laboratory

Rajesh B. Khaparde
Homi Bhabha Centre for Science Education
Tata Institute of Fundamental Research
V. N. Purav Marg, Mankhurd, Mumbai 400088, INDIA

Email: rajeshkhaparde@gmail.com

**Abstract**

The objectives of physics laboratory courses include fostering conceptual understanding and development of several important cognitive, psycho-motor, attitudinal and affective abilities. In most of the Indian colleges and universities (and probably at many other places all over the world) the usual practice of performing a set of experiments, in a 'cook-book' mode, seldom helps students achieve the objectives of the physics laboratory courses and develop the abilities and skills required to become a successful experimental physicist. This paper describes the details of an instructional approach designed and being followed by the author for a past few years, to encourage students' independent thinking in the physics laboratory. This instructional approach encourages students active participation, independent thinking and offers an opportunity to learn 'how to think scientifically' during traditional physics laboratory courses without major 'curriculum' and 'content' changes. Here, 'guided problem solving' approach is adopted by combining 'problem-solving' and 'guided design' modes of instruction. In this approach an experiment is presented as an 'experimental problem' with well thought procedural instructions. During a typical laboratory session, first an introductory demonstration is presented to each group of a few students separately for the first 20 minutes by the mentors as a prelude to the problem and the remaining laboratory time is made available for students to independently 'solve' the experimental problem. The paper also illustrates the approach by describing a familiar experimental problem and the demonstration on 'Electromagnetic damping' with some details of the experimental arrangement, which is being used by the author for training of students and teachers at the introductory university level.

## 1. Introduction

A considerable amount of time is devoted to physics laboratory training in India at the higher secondary, B. Sc. and M. Sc. levels. During a typical physics laboratory course students are made to perform a large number of experiments in a 'cookbook' mode. Most of the experiments performed in teaching laboratories are of 'verification or determination' type. The number of experiments keep on changing but the mode of conducting laboratory sessions more or less remains the same once introduced at the higher secondary till the M. Sc. level.

It is observed that the students are given detailed instructions either orally in the classroom or in the form of written sheets. Teachers/instructors tell too much to the students in order to complete the experiment in the specified time and thereby deprive them of the opportunity to learn by themselves and leaves a very little scope for students' self-planned and independent experimental work. The laboratory training seldom helps a student achieve the objectives of the physics laboratory courses and develop abilities and skills required to become a successful physicist. A large number of teachers and researchers from all over the world have studied and reported on similar concerns and issues related to laboratory courses. These researchers include Barbenza (1972), Boud (1980), Duggan (2002), Gott R (1995), Gott (1999), Hofstein (1982), Khandelwal (1989) and Kruglak (1960).

The laboratory training is an important and indispensible part of teaching of physics and therefore many teachers and researchers have worked on and developed various instructional approaches, experiments and demonstrations. Some major efforts related to the activity based teaching of physics include, Laws (1999), McDermott (1996), Redish (2003), Sokoloff (2004), Sokoloff (2007) and Sokoloff (1997). It is important to note that each of these approaches has its own place, role, importance and problems with the feasibility or use in the regular laboratory training.

In this paper, an attempt towards reformulating existing experiments and demonstrations and development of novel instructional approach for the physics laboratory training is reported. The instructional approach presented here encourages students active participation, independent thinking and offers an opportunity to learn 'how to think scientifically' during traditional physics laboratory courses without major curriculum and 'content' changes. (Khaparde, 2009)

## 2. A novel instructional approach

In this approach 'Guided problem solving' method is adopted by combining 'problem solving' and 'guided design' methods of instruction. Here, an experiment is presented as an 'experimental problem' with guiding procedural instructions. During a typical laboratory session, first an introductory demonstration is presented to each group of a few students separately by the facilitators / mentors for 20 minutes as a prelude to the problem and then remaining laboratory time is made available for students to independently 'solve' the experimental problem. Each demonstration is carefully designed to help students to solve the given experimental problem by introducing either, the basic conceptual understanding required for the problem, the experimental method or technique or the experimental arrangement.

Here, the demonstration serves a specific role of illustration or observation of an event, a concept or principle. The demonstration performed by a teacher or students in a small group in the laboratory help students to recall and refine their conceptual understanding, which students may need to apply in the given experimental problem. Every demonstration, which was designed had a smooth flow of activities, questions, answers, discussion and explanations and was presented in an interactive manner and the interaction between the students and the mentor was triggered through observations, questions and related discussion. The demonstration was presented to stimulate thinking in students and develop cognitive abilities like observation, application, synthesis, interpretation and inferring.

Some key features of the instructional approach are described below:

a) What is an experimental problem ?

A problem is often seen as a stimulus situation for which there is no ready response and the solution calls for either a novel action or a new integration of available actions. Similarly, an experimental problem may be seen as an experimental situation in which one cannot see a ready solution and one needs to perform operations involving combination of conceptual understanding, procedural understanding, scientific processes and skills to arrive at the desired solution.

Here, each experimental problem is presented as a collection of simple smaller experimental stages, which are interdependent or hierarchical in time. In each stage, the students are given simple tasks. The tasks are woven in succession so that the whole problem unfolds through them and students following them are guided stepwise toward the solution, making definite progress through each step. Thus, students solve the experimental problem in graded stages. Each stage may have a different focus, may involve a different type of experimental activities and may aim at different learning outcomes.

b) Guided problem solving

The author employed 'guided problem solving' method in which an experiment was presented as an experimental problem to the students and students were individually given a carefully designed handout for each experimental problem, the corresponding instruction sheet and the answer paper. In this approach, students were guided through the handout to think of and design their own method, to carry out measurements, to analyze data and were thus guided and trained to understand and solve experimental problems. Here, the task or the expected final stage was clearly stated and explained but the detailed instructions, which a student may follow to reach to the desired state were not given. Instead, some starting hints and instructions were given, which may guide the students towards the solution of the problem.

c) Free laboratory ambiance

The approach encouraged a 'free laboratory ambiance where students were made to think about the procedural aspects of experimentation, with a least possible guidance from the teacher. Students were encouraged to carry out self-designed and independent experimental work. The 'free' laboratory ambiance does not refer to an open ended or exploratory type of experimental activities. The idea of 'free' laboratory ambiance was that the students were told about the final outcome of each part of experimental work, but they were given autonomy with respect to, choice of variables, choice of range of values of variables, use of instruments and experimental techniques, method of data handling and analysis etc. For example, in an experimental problem if students are asked to study the relation of incident intensity to the output current of a photodetector, then they may be given a starting instruction on the possible use of inverse square law for establishing linearity, they may be asked to identify the necessary apparatus with their detailed specification, they may be given some hints for the experimental arrangement, and asked to identify the dependent, independent and control variables, construct a fair test, identify the sample size, understand the types of the variables involved and thus in short design the detailed procedure. Then they may be asked to choose sensible values of variables or parameters, proper range and interval between different values of these parameters. They may then be asked to record the desired data and analyze the data using tables and graphs to derive meaningful and expected results. Thus, in this approach the students were provided a freedom with respect to finer procedural stages, but were still guided with respect to the approach or a possible method of solving the problem.

d) Format of presentation

Students were expected to read the student handout and understand the necessary details of the problem. They were expected to understand the use of different apparatus and the related warnings or precautions. Students were expected to broadly use the procedural instructions, design an appropriate method on their own, answer the questions, carryout the necessary measurement, record the data, carryout the necessary analysis of the data and derive the results.

e) Student handout

Students were individually given carefully designed handout for each experimental problem. Students were expected to work independently with guidance provided in graded stages through the handout. The description of the experimental problem given in the handout did not follow 'cookbook' format, instead it took the student away from mechanically followed instructions to a more self designed and student oriented experimental activities. The students were given a brief conceptual introduction to the problem, the necessary description of apparatus and the experimental setup and the theoretical basis through the handout. The students were given the procedural instructions with an intension to 'guide' them but only with respect to a possible method of 'solving' the experimental problem.

f) Procedural instructions

Students were given 'open' procedural instructions, which guided them to a right start and encouraged them to think on various aspects of experimentation. These instructions guided students' thinking at the same time offered a room for independent thinking, designing and planning of actual procedures. These 'open' instructions were not like 'cookbook' type of procedural instructions where students are 'spoon-fed' directly with actual procedural stages without any scope for independent thinking and designing.

For example the instructions included, "You may have to use law of Malus; identify the independent, dependent and control variables; vary the parameter X in convenient and appropriate steps and study its effect on parameter Y; plot an appropriate graph to determine Z; record the necessary data to study the inter-dependence of X and Y; determine the value of X graphically". The instruction "determine the value of X graphically" informed students that they were expected to think of and plot an appropriate graph and determine from it the value of the parameter X, but it did not inform them about what the nature and the scale of the graph should be, which parameters should be plotted, how the parameter X is determined, and so on.

g) Experimental arrangement

Students were provided with the required apparatus and were given a free hand to work in the laboratory. They were also given some extra apparatus and instruments to choose the most appropriate instrument for a particular measurement. Students were supposed to assemble various instruments and make the necessary experimental arrangement on their own.

Students were also given an instruction sheet specifically prepared for a problem. This instruction sheet had information on the use of different instruments and apparatus, the adjustment of the apparatus and the necessary safety instructions and warning. The user manuals of various instruments published by the manufacturers were made available to the students on request.

h) Reporting of laboratory work

Students were not observed by a teacher, while they worked on the given experimental problems and hence were not evaluated on the basis of direct observations by the teacher. Instead, students' performance was only verified or assessed on the basis of the report of their work produced in the answer paper. This aspect of the instructional approach effectively reduces the teachers intervention into the students work and encourages the students self designed independent experimental work. The students' answer paper was treated as the only record of their laboratory work and their solution to the problem. Students were expected to record and report on every procedural step they adopted during the experimental work, observations, readings, method, detailed data analysis, final results and inferences in the answer paper.

i) Preliminary questions

Students were given a set of preliminary questions (often referred to as pre-lab questions) for each experimental problem. Each student was required to independently answer these questions prior to the actual experimental work. These questions were based on the basic concepts, laws, principles, their applications, experimental techniques, use or care of apparatus and a variety of procedural aspects related to the design, measurement and data handling. These preliminary questions played a very important role in preparing the students to efficiently carryout the experimental work.

j) Essentials for experimental physics

It was felt that there are some important aspects related to measurements, statistical treatment of data, graphical representation and analysis of data, significant figures and error analysis, which are essential tools of experimental physics. Students' should have a good knowledge and understanding of these before taking the course in experimental physics. In this approach a detailed reading material on all these aspects was prepared and made available to the students well in advance. Students were expected to read this material and develop the basic knowledge about all the aspects. During the initial stages of the laboratory training, a considerable amount of time was spent on developing students understanding and confidence with respects to use and regular practice of all the essentials for experimental physics.

k) Role of the teacher

Another aspect of the approach is minimal intervention by teachers. Students were not offered any direct advice from the teacher with respect to procedural aspects in solving the experimental problem instead the teacher played a role of a silent observer. The teacher provided minimal guidance to the students. Students were expected and allowed to take their own decisions about procedures and measurements. The teachers were requested to at times coerce students to plan the procedural details individually on their own. Teachers helped students in understanding the use of instruments or even the theoretical basis of the problem.

The approach described above is illustrated below through a set of sample experimental problem and associated demonstration on electromagnetic damping, which the author has developed for the laboratory training of undergraduate physics students.

**3. Sample experimental problem on electromagnetic damping**
In this experimental problem, an aluminum disc is mounted on a horizontal axle around which a cord is wound (Figure 1 and Figure 2). A slotted mass hangs from the free end of the cord. If this slotted mass is released, it accelerate due to the force of gravity. There is a torque on the disc and it undergoes angular acceleration. A pair of cylindrical magnets is placed symmetrically with respect to the disc. If the slotted mass is released in the presence of the magnets, the slotted mass and the disc initially accelerate and soon the disc reaches a constant angular velocity and the slotted mass falls with a terminal velocity. The constant angular velocity of the disc indicates rotational equilibrium of the disc where the torque due to the weight is balanced by an opposite damping torque. There is a frictional torque at the supports, but it is relatively small. The main opposing torque arises due to electromagnetic damping.

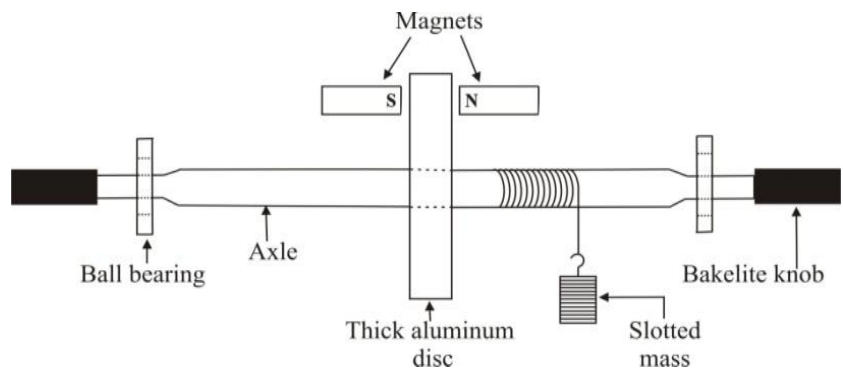

Figure 1. Schematic diagram of disc and magnet assembly

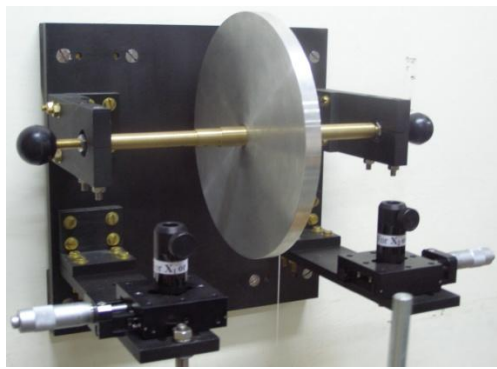

Figure 2. Photograph of disc and magnets assembly

The terminal velocity, with which the weight falls, depends on different parameters like the geometry and dimensions of the disc and the axle, conductivity and magnetic permeability of the material of the disc, the magnitude of the mass attached at the end of the cord, magnetic pole strength of the pair of magnets, position of the magnets, the spacing between the magnets and the disc, and the frictional torque. Two pairs of identical magnets with the same size and shape but of different pole strengths $B_1$ and $B_2$ are given. Also, a velocity measurement unit with a detector was provided to measure the velocity of falling mass.

A student working on this experimental problem was expected to: a) observe the motion of the disc and the falling weights with and without the magnetic field, b) perform necessary measurements to study the variation of the terminal velocity of the falling weight with the magnitude of the mass attached, for a given pair of magnets (with a fixed spacing between each magnet and the disc), c) replace the pair of magnets with another pair, keeping all other parameters the same, and again study and record the variation of the terminal velocity with the mass attached and d) determine the ratio of the magnetic pole strengths of the two pairs of magnets and estimate the frictional torque due to the frictional force at the supports.

**4. Sample demonstration on electromagnetic damping**
The objective of the demonstration was to illustrate the phenomenon of electromagnetic damping and explain the dependence of electromagnetic damping on the conductivity of the material of a conductor in which eddy currents are set up and on the strength of the magnetic field. The experimental setup (Figure 3) consisted of three hollow cylindrical pipes, identical in dimensions and made of aluminum, copper and PVC, respectively. Three small solid (rare earth/ceramic) magnets marked C1, C2, and C3 were used. All the three magnets were identical with respect to their dimensions and masses but had different magnetic pole strengths. The magnet marked C1 was completely de-magnetized by heating it and was not at all a magnet.

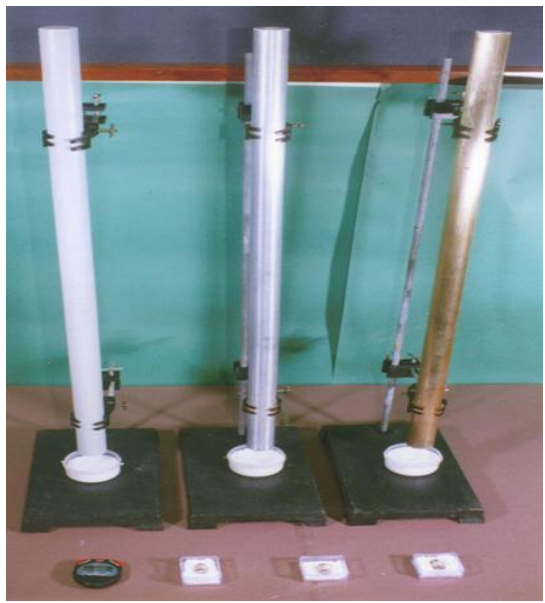

Figure 3. Photograph of the complete experimental setup for the demonstration

In this demonstration one observes the fall of magnets through three different hollow cylindrical pipes made of aluminum, copper and PVC. The motion of the magnets through aluminum and copper pipes is damped on account of electromagnetic damping caused by induced eddy currents in the pipe. The eddy currents are induced due to the motion of the magnets through the pipe. The motion of these magnets through the PVC pipe is un-damped, since no eddy currents are set up in this case. The motion of the completely de-magnetized magnet C1 through all three pipes was un-damped.

**5. Conclusions**

During a laboratory course, the mode in which students work with the given set of experiments is the most important factor in deciding the effectiveness of training in experimental physics. It is felt that there should be a scope for students' self-planned and independent experimental activities during physics laboratory training. The emphasis should be on students' own initiative, on moving them away from 'cookbook' instructions, from spoon-feeding. The approach based on the 'guided problem-solving' with an innovative format of presentation, described above yielded better learning and overall results compared to the traditional method. This approach successfully guides students to think of and design their own method, to carry out measurements, to analyze data and thus imparts training to solve experimental problems. This instructional approach is being used by the author in various programmes for students and teachers in India.

It is noted that this approach helps students for the development of a) Procedural understanding as well as conceptual understanding and practical skills; b) Experimental problem-solving abilities and independent working habits; c) Higher-level cognitive abilities like designing, predicting, observing, classifying, application, synthesis, interpreting and inferring and d) attitudinal aspects and affective abilities like, creativity, curiosity, interest and open-mindedness.

**6. Acknowledgements**

The author is thankful to Profs. Arvind Kumar, H. C. Pradhan and Prof. J. Ramadas for providing the necessary support and facilities for the developmental work and the courses under the Physics Olympiad and the National Initiative on Undergraduate Science (NIUS) programme at HBCSE-TIFR, Mumbai, India.